\documentclass[reprint,superscriptaddress,amsmath,amssymb,aps, prb]{revtex4-1}

\usepackage{graphicx}
\usepackage{dcolumn}
\usepackage{bm}
\usepackage[ruled]{algorithm2e}
\usepackage{multirow,booktabs}
\usepackage{svg}
\usepackage{paralist,siunitx,soul}
\usepackage[version=4]{mhchem}
\usepackage{hyperref}
\hypersetup{colorlinks,urlcolor=black,citecolor=blue,linkcolor=blue}

\begin{document}

\title{Machine-Learning Accelerated Annealing with Fitting-Search Style\\ for Multi-alloy Structure Predictions}
\author{Chuannan Li}
\affiliation{Hefei National Laboratory, Department of Physics, University of Science and Technology of China, Hefei, Anhui 230026, China}
\author{Hanpu Liang}
\affiliation{School of Materials and Physics, China University of Mining and Technology, Xuzhou, Jiangsu 221116, China}
\affiliation{Beijing Computational Science Research Center, Beijing, 100193, China}
\author{Yifeng Duan}
\email{yifeng@cumt.edu.cn}
\affiliation{School of Materials and Physics, China University of Mining and Technology, Xuzhou, Jiangsu 221116, China}
\author{Zijing Lin}
\email{zjlin@ustc.edu.cn}
\affiliation{Hefei National Laboratory, Department of Physics, University of Science and Technology of China, Hefei, Anhui 230026, China}

\begin{abstract}
Structural prediction for the discovery of novel materials is a long sought after goal of computational physics and materials sciences. The success is rather limited for methods such as the simulated annealing method (SA) that require expensive density functional theory (DFT) calculations and follow unintelligent search paths. Here a machine-learning based crystal combinatorial optimization program (CCOP) with a fitting-search style is proposed to drastically improve the efficiency of structural search in SA. CCOP uses a graph neural network energy prediction model to reduce the DFT cost and a deep reinforcement learning algorithm to direct the search path. Tests on six multi-alloys show the energy prediction model is capable of extracting the bonding characteristics of the complex alloys to achieve interpretability. It also achieves high accuracy with a tiny training set (an increment of 30 samples per iteration) by active learning in less than 5 iterations. Comparison with a few conventional methods shows that CCOP finds the lowest-energy structures with the smallest number of search steps. CCOP cuts the computing cost of SA by two orders of magnitude, while providing better search results than SA. CCOP is promising for serving as a broadly applicable tool for the efficient crystal structure predictions.
\end{abstract}

\maketitle

\section{Introduction}

Finding materials with the desired properties is an important task in the condensed matter physics and material sciences, and guides the direction in the structural search field \cite{Review_1,Review_2}.
Alloying is commonly used for obtaining the desired materials, as the structural, electronic, transport and optical properties of alloys can be tuned by varying the compositions \cite{Liang_1, Liang_2, Liang_3}. 
Unfortunately, the complexity of alloy structural search grows exponentially with the number of atoms in a unit cell.
Thus, reducing the computing costs while maintaining the reliability of the search results is of high importance for a structural search method.
To tackle the structural search problem, many algorithms have been developed, such as simulated annealing (SA) \cite{SA_1,SA_2,SA_3}, genetic algorithm (GA) \cite{GA_1,GA_2,GA_3}, particle swarm optimization \cite{PSO1,PSO2}, and \textit{ab initio} random structure search \cite{Random}.
The methods have shown success to various degrees, but there are still large room for improvement. 
For example, the SA and \textit{ab initio} random structure search require many thousands of iterations to reach the low-lying minima in the energy landscape and are expensive due to the CPU demanding density functional theory (DFT) \cite{DFT_1,DFT_2} calculations.
The GA search, however, is extremely dependent on the quantity and diversity of chosen populations. 
Machine learning (ML) is becoming increasingly popular in accelerating the discovery of new materials by encoding physical knowledge into property models \cite{SchNet,MEGNet}. 
For instance, the deep tensor neural network \cite{DTNN} unifies many-body Hamiltonians to design neural network. 
The crystal graph convolutional neural network (CGCNN) \cite{CGCNN} considers the topology of crystal to build graph, providing a universal and interpretable representation of materials.
These methods minimize the need of DFT calculations and have shown high performance in property predictions via the combination of ML and physical concepts. 
Nevertheless, training the ML models to gain the desired generalization capability still requires a large amount of labeled data that often means a large amount of expensive DFT computations. 
Therefore, reducing the number of data required by training the property prediction model (PPM) is a key issue to be resolved for the improved efficiency.
ML may also be used to design the strategy of exploring the potential energy surface (PES) to cut the computational cost of a structural search method. 
As is known, the reward-driven reinforcement learning (RL) focuses on the best policy of exploration in an interactive environment. 
RL has shown success in various fields. 
For instance, AlphaGo \cite{AlphaGo} showed its strong ability in combinatorial optimization to maximize the gain and defeated the world championship in Go game. 
In the fields of physics, chemistry, and biology, RL has been used to design nanophotonic devices \cite{nanophotonics} and drug molecules \cite{drug_1,drug_2,drug_3,drug_4} by learning the policy to optimize the objective function. 
Naturally, RL is expected to be helpful for the structural search methods by learning and making decision on the favorable descent path on PES.
	
In this work, we proposed a crystal combinatorial optimization program (CCOP), which uses a weighted crystal graph convolutional neural network (WCGCNN) as the PPM, and employs a deep reinforcement simulated annealing (DRSA) technique as the searching algorithm. 
The active learning \cite{active_1,active_2,active_3} is applied with the selection of highly representative samples so that the training of PPM requires only a small data set. 
The DRSA is used to guide the search agent to further reduce the computational cost. 
The numerical efficiency of CCOP is illustrated by its applications to the search of the ordered structures of six testing multi-alloys: \ce{BN}, \ce{BeZnO2}, \ce{AlNSiC}, \ce{GaNZnO}, \ce{BeMgCaO3}, and \ce{Al2Ga2Zn2N3O3}.
The test also reveals that the PPM is interpretable, accurate and fast to compute. 
Meanwhile, the DRSA is shown to have the highest performance among the tested search methods, including the conventional RL algorithm of proximal policy optimization (PPO2) \cite{PPO}, SA, and the random search.
\section{Methods}
	
The workflow of CCOP proposed here is illustrated in Fig.~\ref{fig:fig1}. 
It consists of four major parts: i) training set labeling, ii) property prediction model, iii) structural search, and iv) sample selection.
Each part is explained as follows.
	
\begin{figure}[!htb]
	\includegraphics[width=8.3cm]{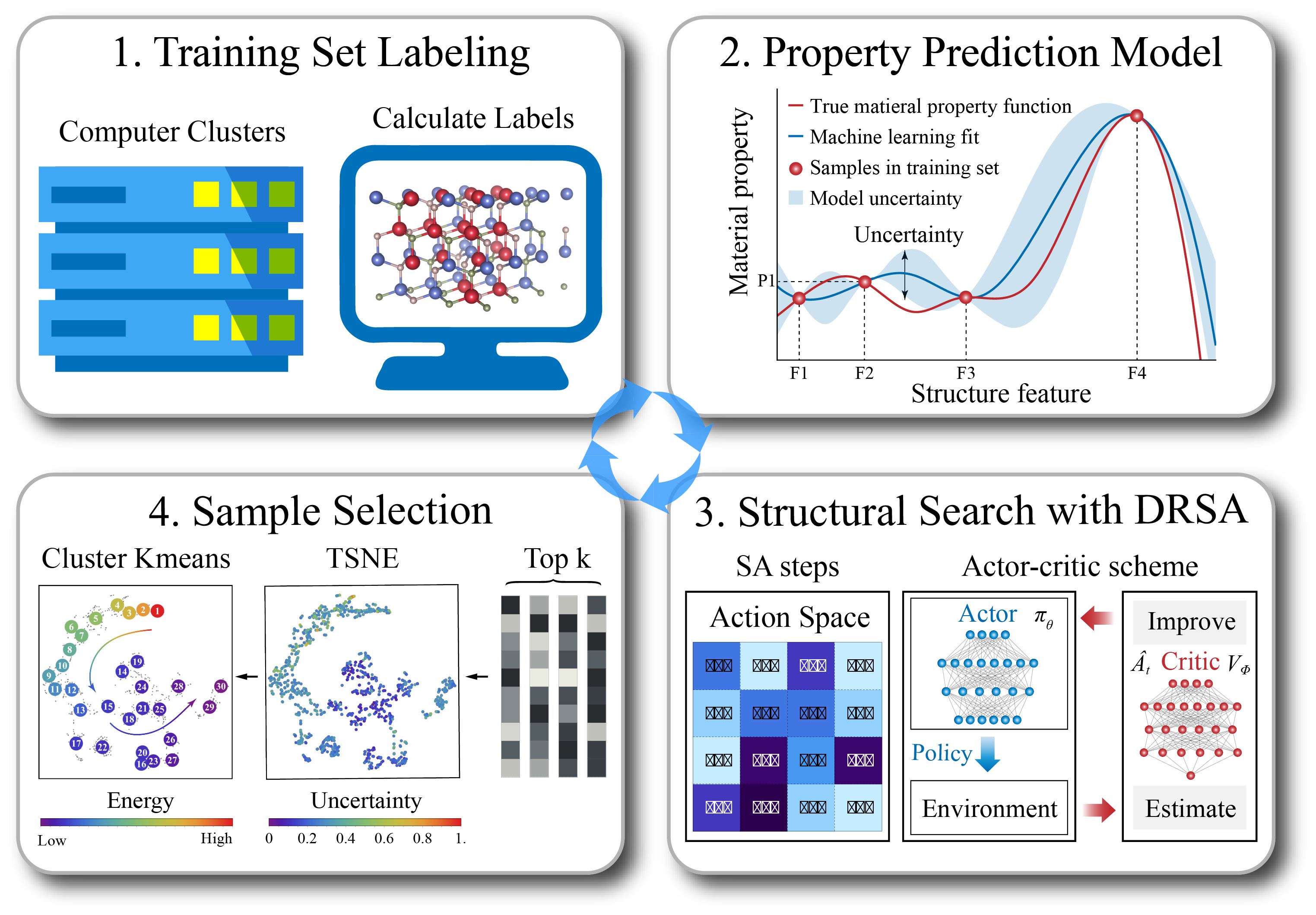}
	\caption{Workflow of the crystal combinatorial optimization program (CCOP). The program starts with a few initial structures as the samples. The samples are labeled with DFT energies and used as the training set (step 1). The labeled training set is used to train a machine-learning potential that fits the PES (step 2). It is followed by training an RL agent for the efficient sampling of the PES to find the low-energy structures (step 3). Then clustering analysis is performed on the sampled structures to select samples for the training set (step 4). The entire optimization program runs in a closed loop.}
	\label{fig:myfig1}
\end{figure}
	
\textbf{Training Set Labeling.}
The training set of structural samples is labeled with the single-point DFT energies that are calculated with the Vienna ab initio Simulation Package (VASP) \cite{VASP_1,VASP_2,VASP_3}. 
The generalized gradient approximation (GGA) with Perdew-Burke-Ernzerhof (PBE) exchange and correlation functional \cite{PBE} is used, and the ion-electron interactions are treated by projector-augmented-wave (PAW) \cite{PAW_1,PAW_2} technique.
The initial training structures are generated by sampling the PES with 30 SA steps. 
30 more structures selected by active learning are added to the training set after every fitting-search iteration of CCOP.
	
\textbf{Property Prediction Model.}
Training a PPM to replace the expensive DFT calculations (step 2 in Fig. \ref{fig:myfig1}) is the second part. The PPM is built based on CGCNN, and modified under the architecture of message passing neural network \cite{MPNN}. 
At the $ k $-th message passing phase, atom vectors $ \bm{h}_{i}^{k} $, $ \bm{h}_{j}^{k} $ and their bond feature vector $ \bm{e}_{ij} $ form the message by function $ M_{k} $. The aggregated message $ \bm{m}_{i}^{k+1}=\sum_{j\in N(i)}M_{k}(\bm{h}_{i}^{k},\bm{h}_{j}^{k},\bm{e}_{ij}) $ is the sum of all $ N(i) $ neighbors of atom vector $ \bm{h}_{i}^{k} $ in crystal graph $ \mathcal{G} $.
New representation of atom vector $ \bm{h}_{i}^{k+1}=U_{k}(\bm{h}_{i}^{k},\bm{m}_{i}^{k+1}) $ is obtained by updating function $ U_{k} $. 
After $ K $ times of message passing, property can be predicted by $ \hat{y}=R(\{\bm{h}_{i}^{K}|i\in \mathcal{G}\}) $, where $ R $ is a differentiable function.
In order to focus on the property related messages in the model, we assign each message with a weight $ \omega_{j}^{k} $ and modify $ M_{k} $ of CGCNN as
\begin{equation}
	M_{k}=w_{j}^{k}\cdot\sigma(\bm{x}_{ij}^{k}\bm{W}_{f}^{k}+\bm{b}_{f}^{k})\odot g(\bm{x}_{ij}^{k}\bm{W}_{s}^{k}+\bm{b}_{s}^{k}),
\end{equation}
where $ \odot $ denotes element-wise multiplication, $ \bm{W}_{f}^{k}, \bm{W}_{s}^{k} $ and $ \bm{b}_{f}^{k}, \bm{b}_{s}^{k} $ are weight matrices and bias vectors of the $ k $-th layer, respectively, and $ \sigma $ is a sigmoid function, $ g $ is a softplus function \cite{softplus}. 
$ \bm{x}_{ij}^{k}=\bm{h}_{i}^{k}\oplus\bm{h}_{j}^{k}\oplus\bm{e}_{ij} $ concatenates neighboring atom pair with their bond. 
The 12 nearest neighboring atoms are found by using a cutoff distance of 8 $ \mathrm{\AA} $. The message weights, $ \omega_{j}^{k} $, for the 12 neighboring atoms are initialized with the same value, since it's hard to tell which message is more important at first.
Moreover, we perform a gated structure \cite{GRU} to control the update process,
\begin{eqnarray}
	U_{k}&&=\bm{z}^{k}_{i}\odot\bm{h}_{i}^{k}+(\bm{I}-\bm{z}^{k}_{i})\odot\bm{m}_{i}^{k+1},\\
	\bm{z}^{k}_{i}&&=\sigma[\bm{W}^{k}_{u}\cdot(\bm{h}_{i}^{k}\oplus\bm{m}_{i}^{k+1})],
\end{eqnarray}
where $ \bm{I} $ is an all-ones vector, $ \bm{W}^{k}_{u} $ is the weight matrix, $ \sigma $ is applied to scale each dimension of $ \bm{z}^{k}_{i} $ in $ [0,1] $,
and weight vector $ \bm{z}^{k}_{i} $ is used to determine the update ratio of $ \bm{h}_{i}^{k} $. The gated structure has shown a good performance in retaining and filtering information \cite{GRU-advantage}.
When the message passing process finish, atom $ i $ is embedded into its chemical environment by iteratively including the surroundings, thus $ \bm{h}_{i}^{K} $ can be treated as the representation of atom $ i $ in the structure. 
As for the representation of crystal structure, we sum up $ \mathcal{N} $ atom vectors and average them as the crystal vector $ \bm{c}=\sum_{i}\bm{h}_{i}^{K}/\mathcal{N} $, which contains machine-learned structural features.
A three-layer fully connected network \cite{Goodfellow} is set as the differentiable function for the property prediction,
\begin{equation}
	\hat{y}=\bm{W}_{3}(g(\bm{W}_{2}(g(\bm{W}_{1}\bm{c}+\bm{b}_{1}))+\bm{b}_{2}))+\bm{b}_{3},\label{equ:mybais}
\end{equation}
where $ \bm{W}s $ are the weights, and $ \bm{b}s $ are the biases. Eq. \ref{equ:mybais} is a universal approximator for nonlinear functions \cite{FFNN}. More details about the WCGCNN based PPM are given in the Section I of the Supplemental Materials (SM).
The PPM is trained with the MXNet \cite{MXNet} framework using the Adam optimizer \cite{Adam} for gradient descent optimization, using the mean square error (MSE) as the loss function (step 2 in Fig. \ref{fig:myfig1}). The training uses 150 samples obtained by SA and DFT calculations as the validation set. Since the size of training set is small and there are more parameters in WCGCNN than CGCNN, it is hard to determine the suitable model in the huge parameter space. Therefore, the weights of the nearest atoms and update ratios are frozen at the beginning of training (60 epochs), namely training the network in a smaller parameter space. The model with the lowest validation loss is then chosen, and the weights and update ratio are unfrozen to fine tune the model (60 epochs). The two-step training technique can ensure the training process goes well.
	
\textbf{Structural Search with DRSA.}
Training an RL agent and use it to direct the SA path for the efficient determination of the lowest-energy structures (step 3 in Fig. \ref{fig:myfig1}) is the third part.
The positions of atoms are encoded as a one-dimensional list, and a sequence of actions are applied to minimize the structural energy. 
An action is defined here as the exchange between atoms with the same sign of electricity, and the exchange between the same atoms is forbidden. 
The allowed actions are weighted by the RL agent and the forbidden actions are masked by 0.
	
At the $ t $-th SA step, the DRSA agent performs action $ a_{t} $ by the $ \epsilon $-greedy policy $ \pi_{\theta}(a_{t}|s_{t}) $ to adjust structure $ s_{t} $ under the Metropolis criterion \cite{SA_1}. 
The value of energy descent $ r_{t+1}=E_{0}-\hat{E}_{t+1} $ is defined as the reward, where $ E_{0} $ is the energy of the search starting structure calculated by DFT and $ \hat{E}_{t+1} $ is the energy of sample predicted by PPM.
The discounted sum of rewards is defined as $ G(\tau)=\sum_{t=0}^{T-1}\gamma^{t}r_{t+1} $, where $ \tau=\{s_{0},a_{0},s_{1},r_{1},a_{1},\cdots,s_{T},r_{T}\} $ is a trajectory of Markov decision process and $ \gamma $ is the discount factor determining the priority of short-term rewards. 
Minimizing the structural energy within $ T $ steps is equivalent to maximizing the expectation of $ G $.
	
The agent's policy $ \pi_{\theta}(a_{t}|s_{t}) $ is a key for finding the optimal structure.
The agent is trained by the clipped loss of PPO2 \cite{PPO},
\begin{equation}
	\mathcal{L}_{\mathrm{A}}=-\hat{\mathbb{E}}_{\tau,t}[\min(p_{t}(\theta)\hat{A}_{t},
	\mathrm{clip}(p_{t}(\theta),1-\epsilon,1+\epsilon)\hat{A}_{t})], \label{eq:myloss}
\end{equation}
where $ p_{t}(\theta)=\pi_{\theta}(a_{t}|s_{t})/\pi_{\theta_{old}}(a_{t}|s_{t}) $ is the probability ratio of current and old policies, $ \hat{A}_{t} $ is the estimator of advantage \cite{GAE}. The $ \mathrm{clip} $ function in Eq. \ref{eq:myloss} removes the incentive of $ p_{t} $ beyond the interval $ [1-\epsilon,1+\epsilon] $, and means a penalty for a large policy update.
To reduce the variance of $ \hat{A}_{t} $, the TD(0) \cite{Sutton} form of $ \hat{A}_{t} $ is adopted, i.e., $ \hat{A}_{t}=r_{t+1}+\gamma V_{\pi}(s_{t+1})-V_{\pi}(s_{t}) $, where the state-value function $ V_{\pi}(s_{t})=\mathbb{E}_{\tau}[r_{t}|s_{t}] $ is the expected return from state $ s_{t} $. $ V_{\pi}(s_{t}) $ can be approximated by minimizing the loss \cite{A3C}
\begin{equation}
	\mathcal{L}_{\mathrm{C}}=\hat{\mathbb{E}}_{\tau,t}[r_{t+1}+\gamma V_{\pi}(s_{t+1})-V_{\pi}(s_{t})]^{2}.
\end{equation}
Through learning from the search paths, the agent's policy $ \pi_{\theta}(a_{t}|s_{t}) $ generates a suitable weight for each action to minimize energy, thus improving the search efficiency of SA actions.
More information about the DRSA can be found in the Section II of SM. 
\textbf{Sample Selection.}
Choosing the representative samples to add to the training set (step 4 in Fig. \ref{fig:myfig1}) is the last part of the workflow.
To be most beneficial for reducing the mean absolute error (MAE) of PPM predictions, samples with the highest uncertainties should be considered.
The uncertainty $ \Omega $ is defined as the variance of predictions \cite{active_1}
\begin{equation}
	\Omega(\bm{x})=\dfrac{1}{M-1}\sum_{m=1}^{M}(f_{m}(\bm{x})-\dfrac{1}{M}\sum_{l=1}^{M}f_{l}(\bm{x}))^{2}, \label{eq:myomega}
\end{equation}
where $ \bm{x} $ is a searched sample, $ f_{m} $ denotes a trained PPM, and $ M $ is the number of PPMs.
Specifically, 10\% of the samples with the highest uncertainties and 10\% of the samples with the lowest energies are selected. The crystal vectors of these samples are calculated by the PPM, followed by $ t $-distributed stochastic neighbor embedding (TSNE) \cite{TSNE} to reduce the dimension of crystal vectors. 
The reduced vectors are grouped into 30 clusters by the Kmeans method \cite{Kmeans}. 
The minimal energy sample in each cluster with its DFT energy computed is added into the training set. 
Moreover, the lowest energy sample in the training set, with the energy referred as $ E_{0} $ above, is used as the initial structure of the next fitting-search iteration.
\section{Results and Discussion}
	
The effectiveness of the CCOP is tested by searching the lowest-energy structure of the ordered configurations of multi-alloy. 
Six alloys with compositions from simple to complex are used as the testing cases: \ce{BN}, \ce{BeZnO2}, \ce{AlNSiC}, \ce{GaNZnO}, \ce{BeMgCaO3}, and \ce{Al2Ga2Zn2N3O3}.
The search is restricted to that the alloys maintain the 72-atom wurtzite-like lattice configuration in any temperature environment \cite{BN, BeZnO2, GaNZnO}.
The unit cell consists of 8 lattice layers and there are 9 atomic sites in each layer. Initially each layer is filled with 9 cations or anions, and the cation-layer and the anion-layer alternate.
The atomic arrangements are then changed in the search process to obtain lower energy configurations.
As there is only one type of anions (cations) in BN, the search action in BN refers to the exchange of the positions of anion and cation, instead of exchanging among cations or among anions for the structural searches of the other five alloys.
	
\subsection{Model Interpretability}
	
\begin{figure*}[!htb]
	\centering
	\includegraphics[width=16cm]{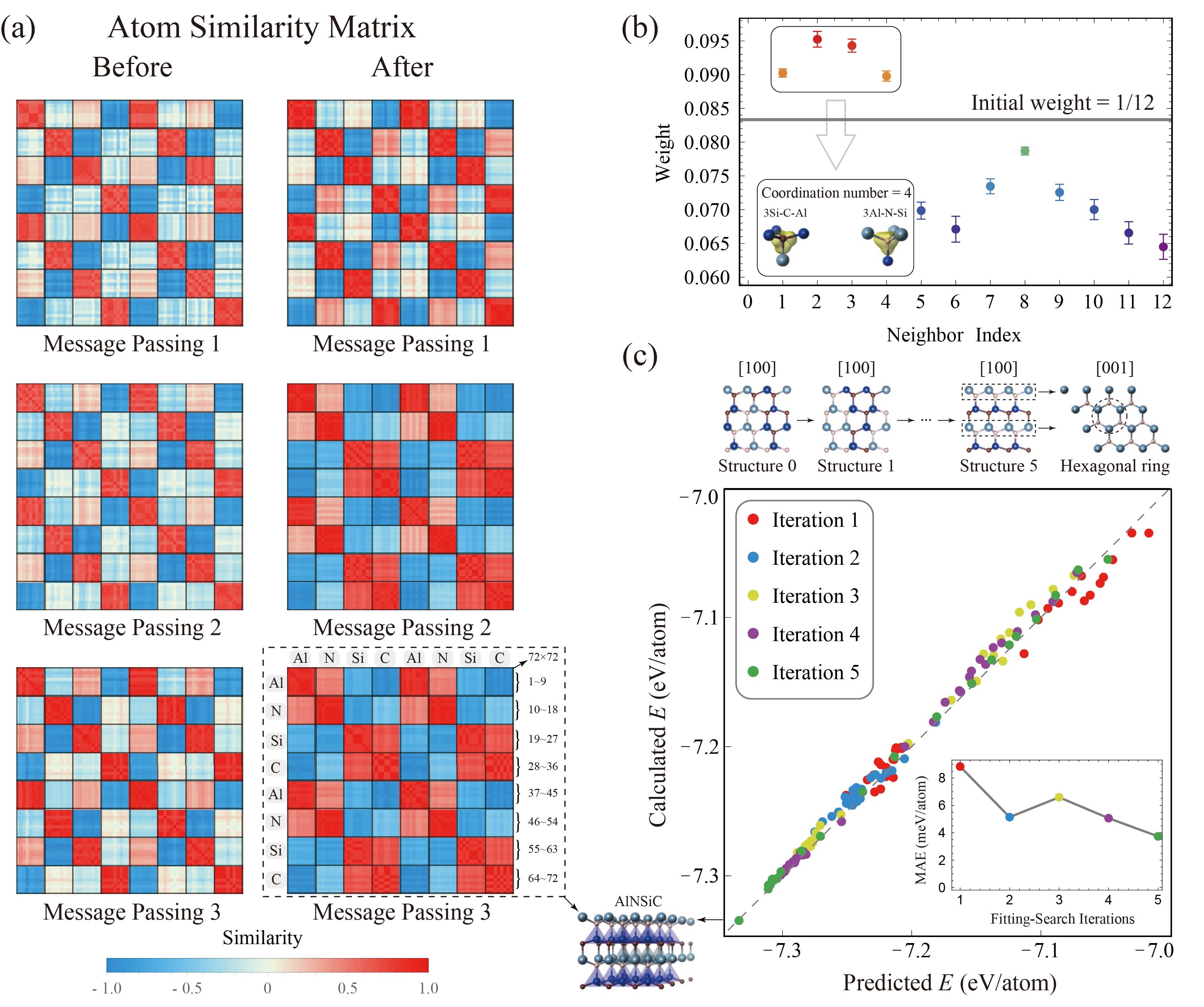}
	\caption{(a) The atom similarity matrix of the ordered AlNSiC structure before and after PPM training. The similarity coefficient is defined as the cosine distance between the atom vectors. (b) Atomic weights of the 12 neighboring atoms by the trained PPM, notice that the nearest 4 atoms have higher weights than the others. (c) Comparison between the PPM predicted energies and the calculated DFT energies, with the inset showing the mean absolute error (MAE) of the predictions.}
	\label{fig:myfig2}
\end{figure*}
	
Interpretability of ML model, e.g., the characteristic vectors extracted from the crystal information are consistent with the physical intuition, is desired in the fields of physics, chemistry and material sciences. 
Here, AlNSiC is used as an example to show the interpretability of our PPM. 
We label each atom vector from 1 to 72 and each layer from 1 to 9, then construct the atom similarity matrix, with a size of $72\times 72$, under different message passing phases. 
As seen in Fig. \ref{fig:myfig2}(a), the distribution of the atom similarity matrix before the PPM training mainly depends on the input atom features, e.g., electronegativity and valence electrons. 
The atom similarity matrix is almost unchanged after three message passing process, which indicates that the PPM cannot extract effective information from the input structures.
After the training, atoms are gradually separated into two clusters, i.e., the Al-N and Si-C clusters. 
For example, the Al atoms (1$\sim$9) show a higher similarity with the N atoms (10$\sim$18) than the Si atoms (19$\sim$27) and C atoms (28$\sim$36), consistent with the fact that Al bonds with N, not Si or C.
	
The results are consistent with the structural characteristics of wurtzite structure and the matching of valence electrons.
For instance, the binary compound AlN forms a stable wurtzite structure by the $sp^3$ hybridization for $3s^23p$ and $2s^22p^3$ electrons of Al and N atoms, respectively. 
Similarly, SiC forms a wurtzite structure by $sp^3$ hybridization for $3s^23p^2$ electrons of Si and $2s^22p^2$ electrons of C.
Thereby the Al atoms at the lowest-energy order structure must be bonding with N first, while the high-energy structures have more random atomic distributions, as seen on the top of Fig. \ref{fig:myfig2}(c).
Moreover, the Al-N and Si-C layers, which form the hexagonal rings with the neighbor atoms along the [001] direction, usually appear alternatively in the lowest-energy structure due to the crystal periodicity.
Thus the Al and N atoms show high similarity in the atom similarity matrix, corresponding to the strong tendency of their bonding.
	
\begin{figure*}[!htb]
	\centering
	\includegraphics[width=17cm]{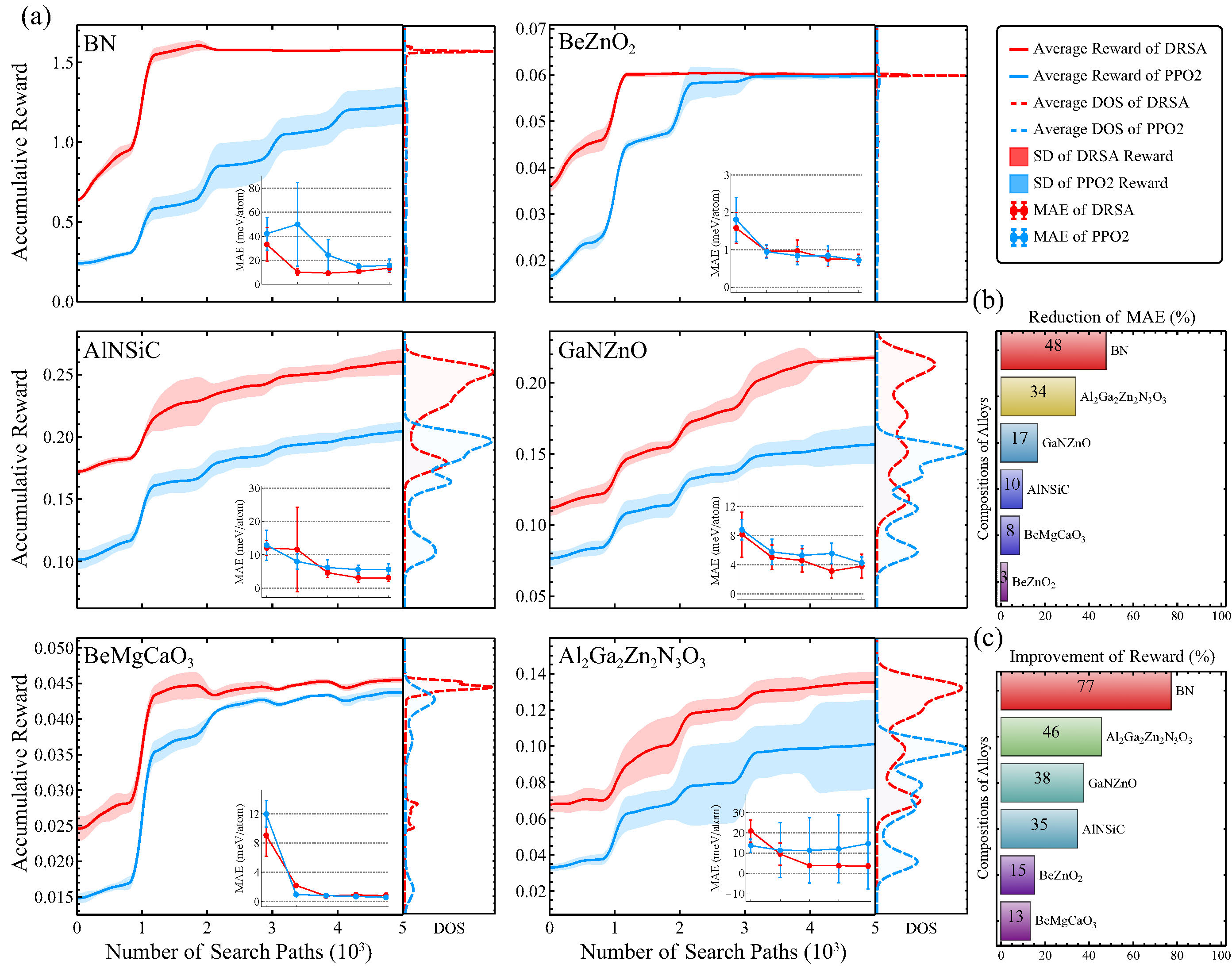}
	\caption{ Performances of DRSA and PPO2 on searching the low energy structures of 6 multi-alloys: (a) The accumulative reward (AR, in eV) and the density of state (DOS) of DRSA and PPO2 versus the search paths. The shadow of the line is the standard deviation (SD) of the ARs. The inner panel shows the change of MAE with the fitting-search iteration. (b) Average reduction of MAE by DRSA relative to PPO2, and (c) Improvement on the AR by DRSA relative to PPO2.}
	\label{fig:myfig3}
\end{figure*}
	
Fig. \ref{fig:myfig2}(b) shows that, although the initial weights are uniformly set to 1/12, after the training, the weights of the 4 nearest neighbor atoms become larger, while the weights for the other 8 neighbors become smaller. That is, the energy prediction is predominately determined by the 4 nearest neighbor, and it matches perfectly with the four-coordinated tetrahedrons, e.g., 3Si-C-Al and 3Al-N-Si, in the lowest-energy structure of AlNSiC.
The atom similarity matrix and the weights verify that the PPM can effectively extract the structural characteristics from the training data. 
The learned weights ensures that the choice of atom exchanges is not as random as the SA, and reduces the cost of choosing the energy descent path in DRSA.
	
Fig. \ref{fig:myfig2}(c) displays the PPM predicted energies against the DFT calculated values and the mean absolute error (MAE) for each fitting-search iteration. 
Based on the active learning, 30 most representative samples from the DRSA paths in each fitting-search iteration are selected to enhance as much as possible the prediction accuracy of PPM. 
As seen in Fig. \ref{fig:myfig2}(c), with the progress of the fitting-search iteration, the searching area gradually moves from the initial high-energy structures to the low-energy structures, and the corresponding MAE usually decreases. 
The results reflect that the search program is capable of effectively finding the low-energy area in the PES and finally obtains the ordered structure of AlNSiC with an energy of -7.33 eV/atom. 
Notice that all the energy values shown below refer to the energies per atom.
	
The interpretability of ML model as shown in Fig. \ref{fig:myfig2} is a common feature of our PPM for all the alloys tested. Additional example can be found in Section III of the SM.

\begin{figure*}[!htb]
	\includegraphics[width=16.5cm]{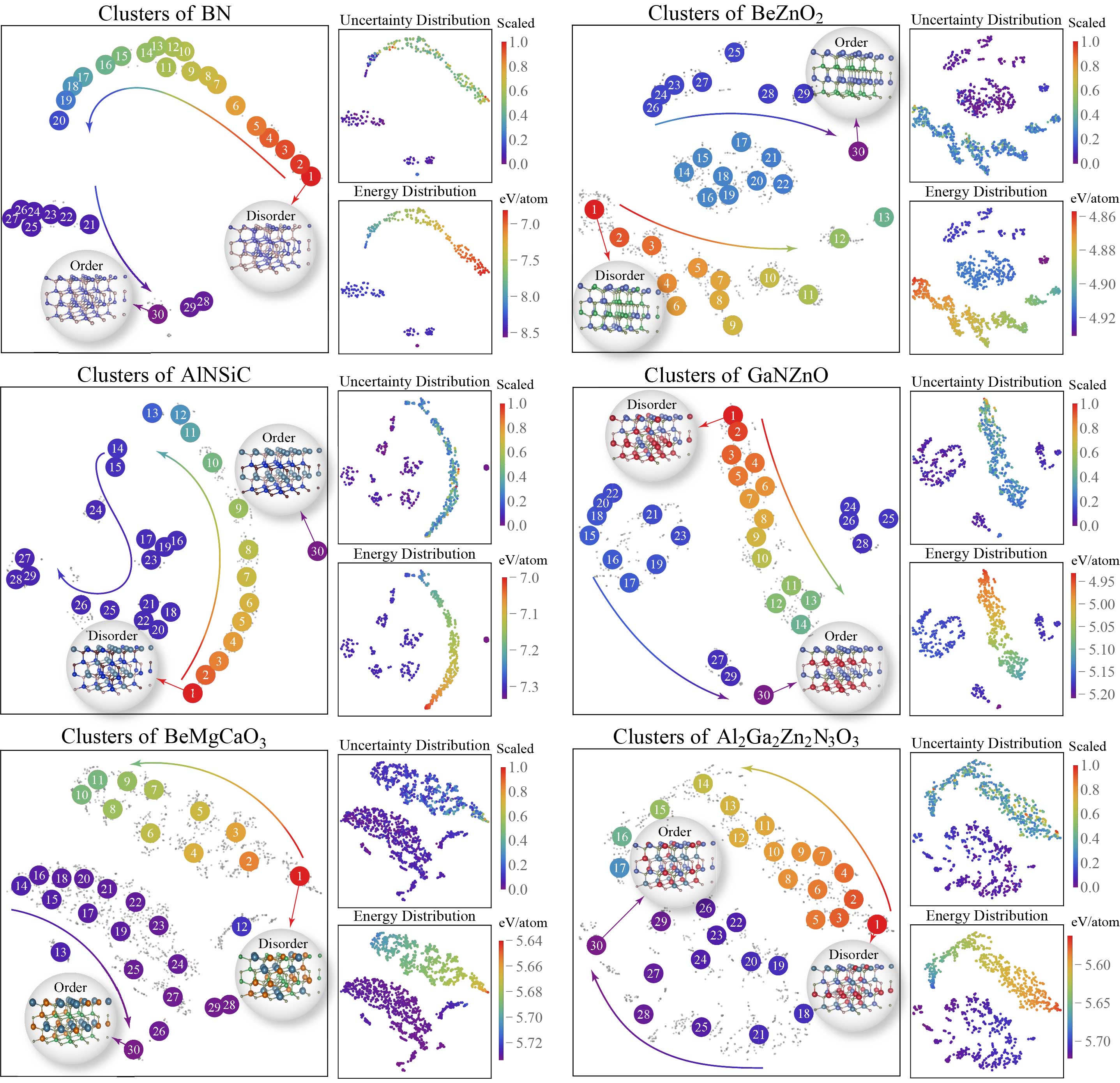}
	\caption{ The distribution of the clusters, the uncertainties, and the energies of selected samples for the six multi-alloys at the last fitting-search iteration. The $x$- and $y$-axis are the feature vectors generated by TSNE. Each cluster is labeled and colored by the rank of energy. The arrow curves show the direction of energy descent.} 
	\label{fig:myfig4}
\end{figure*}

\subsection{Search Ability of DRSA and PPO2}
	
Fig. \ref{fig:myfig3} compares the performances of DRSA and PPO2 on searching the low energy structures of 6 multi-alloys, averaged over 5 separate DRSA or PPO2 runs for every alloy. 
Each of the 5 DRSA or PPO2 runs consists of 5 fitting-search iterations and 1000 search paths per iteration, for a total of 5000 search paths per run. 
Fig. \ref{fig:myfig3}(a) shows the accumulative rewards (ARs) of DRSA and PPO2 for searching the low energy structures of 6 alloys. 
AR is calculated as the total energy descent $E_{0}-\hat{E}_{n}$, where $ \hat{E}_{n} $ is the minimum predicted energy at the $ n $-th search path. 
As seen in Fig. \ref{fig:myfig3}(a), AR of DRSA is higher than that of PPO2 for every alloy examined. 
The results are understandable as DRSA is basically PPO2, but with the policy complexity reduced by the physical constraint of SA. 
With the help of the PPM agent that learns the weight of atomic exchange, which ensures the choice of atom exchanges is not as random as the SA, DRSA requires no more than 5000 search paths in reaching its maximum reward for most alloys. 
The fast convergence is achieved also because that the PPO2 agent learns the energy descent policy from the paths of DRSA and the fitting-search style relieves the difficulty of training the agent, resulting in a reduced number of steps to reach the maximal reward.
	
\begin{table}[tp]
	\caption{Number of executable actions (Actions), number of possible structures (Structures), and the improvement of sample efficiency (ISE) by DRSA relative to PPO2 for six multi-alloys.}
	\label{tab:mytable1}
	\begin{ruledtabular}
		\renewcommand{\arraystretch}{1.1}
		\begin{tabular}{lccc}
			Type of alloys & Actions & Structures & ISE (\%) \\ \hline \rule{-3pt}{10pt}
			$ \mathrm{BN} $ & 1296 & $ 10^{20} $ & 1370\\
			$ \mathrm{BeZnO_{2}} $ & 324 & $ 10^{9} $ & 198\\
			$ \mathrm{AlNSiC} $ & 648 & $ 10^{19} $ & 349\\
			$ \mathrm{GaNZnO} $ & 648 & $ 10^{19} $ & 320\\
			$ \mathrm{BeMgCaO_{3}} $ & 432 & $ 10^{15} $ & 135\\
			$ \mathrm{Al_{2}Ga_{2}Zn_{2}N_{3}O_{3}} $ & 756 & $ 10^{25} $ & 192
		\end{tabular}
	\end{ruledtabular}
\end{table}
	
DRSA uses fewer samples and performs better than PPO2.
The improvement may be measured with the improvement of sample efficiency (ISE) defined as $ \mathrm{ISE}=N_{\text{PPO2}}/N_{\text{DRSA}}*100\% $, where $ N_{\text{PPO2}} $ and $ N_{\text{DRSA}} $ are the number of samples searched by PPO2 and DRSA, respectively. 
ISE, together with the number of executable actions (allowed exchanges of atom positions) and the number of possible structures for each of the six alloys are shown in Table \ref{tab:mytable1}. 
The number of possible structures, or the possible permutations of the atom sites, is found to be from $10^9$ to $10^{25}$ for the 6 alloys. 
As the number of possible structures is very large, certainly vastly larger than $ N_{\text{PPO2}} $ and $ N_{\text{DRSA}} $, employing search methods such as DRSA and PPO2 is necessary in practice. 
Meanwhile, DRSA is superior with a smaller number of samples and a larger AR than PPO2.

Among the 6 alloys tested, the binary alloy BN shows the highest ISE of about 1370\%. 
This may be attributed to that BN has the largest number of executable actions, 1296, at each search step (see Table \ref{tab:mytable1}). 
In fact, without the constraint of SA, the random search at the beginning of PPO2 causes an inefficient exploration of the low-energy area. 
PPO2 also shows a slow convergence of the policy in a large action space. 
The lowest-energy structure is often missing even with 5 different runs of PPO2 (see Table S2 of SM for more testing results). When the number of actions is relatively small, however, the difficulty of policy learning for the RL agent is much reduced and the performance of PPO2 is much improved. 
As a result, PPO2 may catch up DRSA on the accumulated reward, as shown in Fig. \ref{fig:myfig3} for \ce{BeZnO2} and \ce{BeMgCaO3} with the number of actions of 324 and 432, respectively.
The MAE of the predicted energies in each fitting-search iteration reflects the variation of the accuracy of PPM. 
As seen in the insets of Fig. \ref{fig:myfig3}(a), the MAE for either DRSA or PPO2 normally decreases at first due to the increase in the training data, and then converges because of the lack of diversity in the newly added representative samples. Overall, however, DRSA not only shows higher ARs, but also lower MAEs than PPO2 for the 6 tested alloys. 
To be more quantitative, Fig. \ref{fig:myfig3}(b)(c) show respectively the reduction of MAE and the improvement of reward by DRSA relative to PPO2 for the tested alloys. Here, the reduction of MAE is calculated as $ \sum_{i=1}^{5}=(\mathrm{MAE}^{\mathrm{PPO2}}_{i}-\mathrm{MAE}^{\mathrm{DRSA}}_{i})/\mathrm{MAE}_{i}^{\mathrm{PPO2}}*100\% $, where $ \mathrm{MAE}_{i}^{\mathrm{DRSA}} (\mathrm{MAE}_{i}^{\mathrm{PPO2}})$ is the MAE of DRSA (PPO2) at the i-th iteration. The improvement of reward is calculated as $ \sum_{i=1}^{5000}(\mathrm{AR}_{i}^{\mathrm{DRSA}}-\mathrm{AR}_{i}^{\mathrm{PPO2}})/\mathrm{AR}_{i}^{\mathrm{PPO2}}*100\% $, where $ \mathrm{AR}_{i}^{\mathrm{DRSA}} (\mathrm{AR}_{i}^{\mathrm{PPO2}}) $ is the AR of DRSA (PPO2) at the i-th search path.

As can be seen in Fig. \ref{fig:myfig3}(b)(c), there is a positive correlation between the magnitudes of the reduction of MAE and the improvement of reward, both are the largest for BN and the second largest for \ce{Al2Ga2Zn2N3O3}. 
While BN has the largest number of executable actions, \ce{Al2Ga2Zn2N3O3} has the largest number of possible structures and the second largest number of executable actions (see Table \ref{tab:mytable1}). 
Combined with a complicated chemical composition, \ce{Al2Ga2Zn2N3O3} is hard for PPO2 to handle. Not only the SD of AR increases with the search, the SD of MAE (see the error bar of MAE) also increases, from 3.3 meV/atom for the 1st iteration to 22 meV/atom for the 5th iteration. 
Hence, PPO2 is unstable when applied to \ce{Al2Ga2Zn2N3O3}. 
Detailed data analysis shows that, due to a lack of Metropolis criterion to constrain the search direction, a significant number of high energy structures are produced by PPO2, causing difficulty in training PPM. 
It then leads to high MAE of PPM and improper exploration of PES. On the contrary, a positive feedback between fitting and search is formed in the constraint search of DRSA, leading to a continuous reduction of MAE and a stable search.

\subsection{Clustering Analysis}
	
To illustrate the benefit of the active learning for the sample selection, the results of the clustering analysis at the last fitting-search iteration are displayed in Fig. \ref{fig:myfig4}.
As shown in Fig. \ref{fig:myfig4}, the structural samples are grouped by Kmeans into 30 clusters, with the minimal sample energy decreases from cluster 1 to cluster 30 and the energies are similar for samples in the same cluster. 
The uncertainty of the PPM predictions (Eq. \ref{eq:myomega}) decreases with the decreased sample/cluster energy, indicating that the PPM is adaptive during the search, especially to the low-energy area of the PES. 
The correlation between the sample/cluster energies and uncertainties may be quantified with the Pearson correlation coefficient. As shown in Fig. S4 of the SM, the Pearson correlation coefficients are found to be higher than 0.9 for all 6 alloys, with the highest of 0.951 for \ce{Al2Ga2Zn2N3O3}, further illustrating the adaptability of PPM.
	
\begin{figure*}[!t]
	\includegraphics[width=16.5cm]{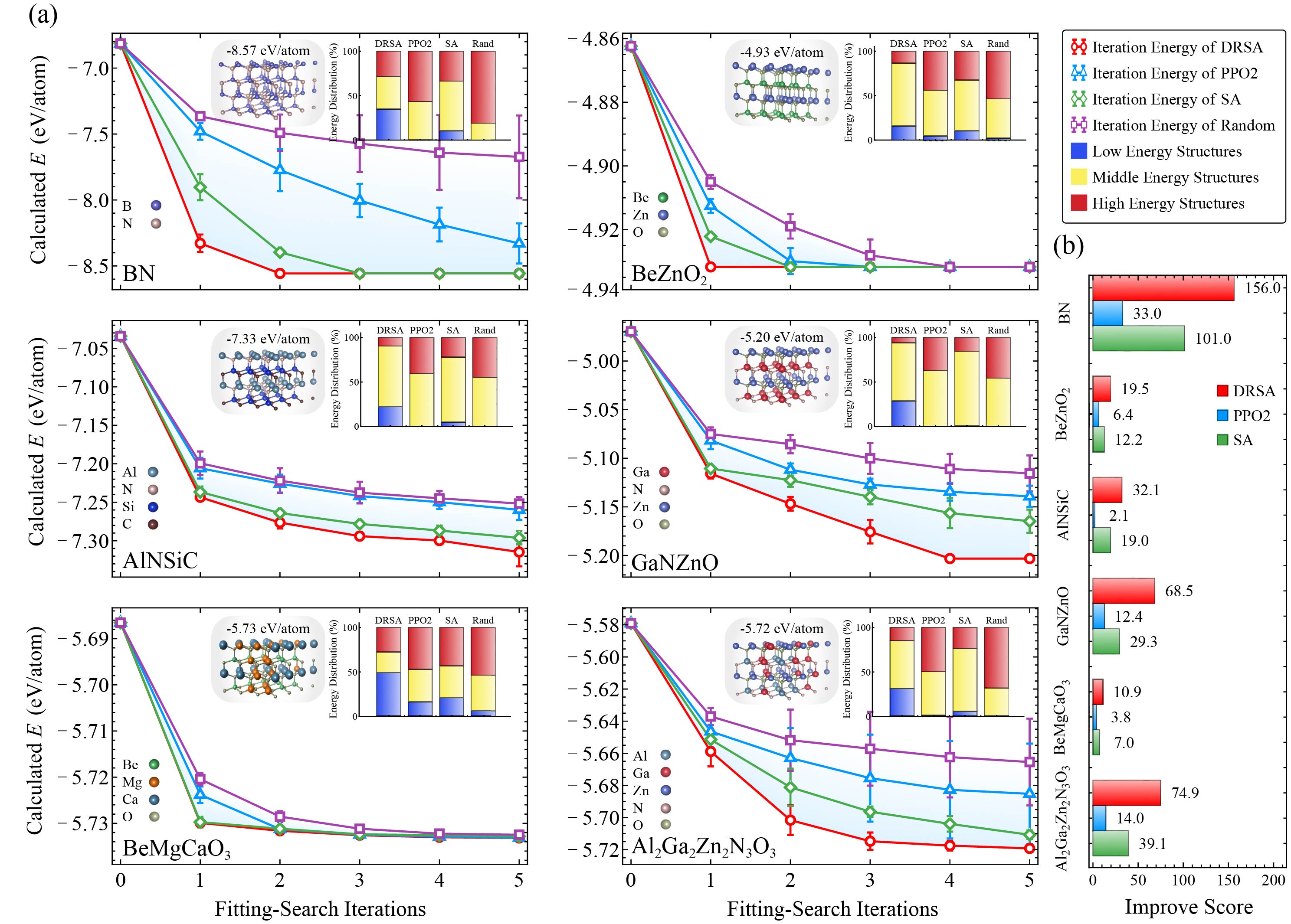}
	\caption{(a) The calculated energy versus the fitting-search iteration for the six multi-alloys. The inner panels are the lowest-energy structures and the energy distribution percentage of the search results. (b) The improvement scores of DRSA, PPO2, and SA for the six multi-alloys.}
	\label{fig:myfig5}
\end{figure*}
	
The proximity of the energies for structures in the same cluster and the strong correlation between the energies and uncertainties are the results of the feature classification of the PPM. 
Therefore, adding only the lowest-energy structure of each cluster to the training set is quite reasonable and has the benefit of minimizing the size of the training set. 
Meanwhile, adding the the high energy structures with high PPM prediction uncertainties, as mentioned in the sample selection strategy above, can help the PPM to fit the entire PES better. 
Consequently, the active learning in the sample selection has a positive effect on the structural searching. 
It improves the accuracy of PPM by putting the most representative samples on the energy landscape in the training set. 
It also reduces the times of expensive DFT calculations \cite{Active_DFT_1,Active_DFT_2} effectively by reducing the size of the training set.
The DRSA predictions are also coherent with the physical intuition. 
The lowest-energy structures of \ce{AlNSiC}, \ce{BeZnO2}, \ce{BN}, and \ce{GaNZnO} are predicted to be ordered structures with the obvious layered characteristics. 
Specifically, \ce{GaNZnO} consists of the Ga-N and Zn-O layers placed alternatively along the $z$-axis direction (see the insets in Fig. \ref{fig:myfig4} ), providing the best match of the valence electrons for the stabilization of the $sp^3$ hybridization.
For the quaternary alloy \ce{BeMgCaO_{3}}, the Be and Ca atoms are placed in different layers due to the large difference in their atomic sizes, while the Mg atoms are evenly distributed in the Be- and Ca-layer. 
Similar structural feature is observed in the lowest energy structure of the quintary alloy \ce{Al2Ga2Zn2N3O3}. 
\subsection{Method Comparison} 
	
The lowest energies and the structural energy distributions of the 6 alloys versus the searching iterations for the methods of DRSA, PPO2, SA, and the random search are shown in Fig. \ref{fig:myfig5}(a). The performance of different methods may be measured by the improvement score which is defined as $ s = p\cdot\dfrac{1}{5}\sum_{i=1}^{5}|\Delta E_{i}|/|\Delta E_{i}^{R}|\cdot 100\% $.
Here $p$ is the weighted average concerning the structural energy distribution, with the weight ratios of 6:3:1 for the low-, middle-, and high-energy structures. $|\Delta E_{i}|=|E_{i}-E_{0}|$ is the energy difference between the lowest-energy structure of the $ i $-th iteration and the initial structure, and we use the random search $ |\Delta E_{i}^{R}| $ as the benchmark.
The improvement scores of DRSA, PPO2, and SA for the 6 alloys are shown in Fig. \ref{fig:myfig5}(b). 
On average, the improvement scores of DRSA, PPO2, and SA for the 6 alloys are 60.37\%, 11.93\% and 34.62\%, respectively. 
DRSA has the highest score due to its combination of RL and SA, while PPO2 has the lowest performance due to its inefficiency in a large action space. More information about the performances of the 4 search methods can be found in Table S2 of SM. 
As seen in Table S2, DRSA usually finds the lowest-energy structure with the smallest number of iterations. Moreover, all the 6 lowest-energy structures, one for each of the 6 alloys, are found by DRSA. However, 4 of them are missed by PPO2 and SA, while 5 of them are missed by the random search. 
Clearly, the newly proposed method of CCOP that combines the WCGCNN, SA, and the RL is capable of drastically reducing the computational cost, while maintaining the desirable accuracy. Compared to SA that is more efficient than PPO2 and the random search (Table S2), CCOP may reduce the computational cost from 7$ \sim $9 days for SA to 1$ \sim $2 hours (Section V of SM), i.e., a reduction of two orders of magnitude.
	
\section{Conclusions}
	
A machine-learning assisted structural prediction method named as CCOP is proposed. The novel features of CCOP include: i) Using a modified CGCNN model as the PPM to replace the expensive DFT calculations. ii) Guiding the structural search paths by DRSA, a method combining the advantages of RL agent and Metropolis criterion, to accelerate the searching process. iii) Employing an active learning based sample selection method to reduce the PPM prediction error and minimize the size of the training set.
	
Through testing applications concerning the structural searches of 6 multi-alloys, it is demonstrated that: i) The PPM has the desired feature of interpretability, since the results of the atom similarity matrix and the atom exchange weights for the nearest neighbors are both consistent with the physical intuition. ii) DRSA outperforms SA, PPO2 and the random search approach by finding the lowest energy structures with the smallest number of steps. The benefit of DRSA is more pronounced when considering that SA, PPO2 and the random search miss most of the lowest-energy structures for 6 alloys. iii) Selecting samples through active learning makes the PPM adaptive during the search, resulting in an efficient exploration of the low-energy area of the PES. 

Overall, the integrated search framework of CCOP is found to cut the computational cost of a conventional SA by two orders of magnitude. CCOP should be useful for the speedy discovery of novel materials.

\begin{acknowledgments}
	The work is sponsored by the National Natural Science Foundation of China (Nos. 12074362, 11774416 \& 11774324).
	We thank the USTC supercomputing center for providing computational resources for this project.
	
	C.L. and H.L. contribute equally to this work.
\end{acknowledgments}

\bibliographystyle{apsrev4-1}

%

\end{document}